\documentclass[runningheads]{llncs}

\usepackage[colorlinks, allcolors=blue]{hyperref}
\usepackage{amsmath,amssymb,amsfonts}
\usepackage{multirow}
\usepackage{utfsym}

\usepackage[misc]{ifsym}

\usepackage[T1]{fontenc}
\usepackage{graphicx}
\usepackage{color}

\begin{document}
\title{BeeTLe: An Imbalance-Aware Deep Sequence Model for Linear B-Cell Epitope Prediction and Classification with Logit-Adjusted Losses}
\titlerunning{A Framework for B-Cell Epitope Prediction}
%
\author{Xiao Yuan [\Letter]}
\authorrunning{X. Yuan}
%
\institute{Georgia Institute of Technology, Atlanta, USA \\
\email{xyuan8@gatech.edu}}

\tocauthor{Xiao Yuan}
\toctitle{BeeTLe: A Framework for Linear B-Cell Epitope Prediction and Classification}

\maketitle              
\begin{abstract}
The process of identifying and characterizing B-cell epitopes, which are the portions of antigens recognized by antibodies, is important for our understanding of the immune system, and for many applications including vaccine development, therapeutics, and diagnostics. Computational epitope prediction is challenging yet rewarding as it significantly reduces the time and cost of laboratory work. Most of the existing tools do not have satisfactory performance and only discriminate epitopes from non-epitopes. This paper presents a new deep learning-based multi-task framework for linear B-cell epitope prediction as well as antibody type-specific epitope classification. Specifically, a sequenced-based neural network model using recurrent layers and Transformer blocks is developed. We propose an amino acid encoding method based on eigen decomposition to help the model learn the representations of epitopes. We introduce modifications to standard cross-entropy loss functions by extending a logit adjustment technique to cope with the class imbalance. Experimental results on data curated from the largest public epitope database demonstrate the validity of the proposed methods and the superior performance compared to competing ones.

\end{abstract}
\section{Introduction}
In our adaptive immune system, B cells play a critical role by producing antibodies that detect, neutralize and help eliminate the pathogens, such as viruses. Antibodies can recognize and bind to antigens, which are usually proteins, on the pathogens. These bound regions are called epitopes and they can be divided into linear and conformational epitopes. Although the majority of the B-cell epitopes are conformational, much attention is concentrated on the identification of linear epitopes, which consist of a contiguous sequence of amino acids (residues). The reason is that linear epitopes can be used to design peptide-based vaccines and replace infectious antigens in antibody production and diagnostic assay development~\cite{potocnakova2016introduction}. Since experimental epitope mapping is time-consuming, costly, and laborious, computational prediction methods are desirable to reduce the number of potential epitope candidates for experimental validation~\cite{sanchez2017fundamentals}.

With the ever-increasing data of verified epitopes, machine learning-based approaches are developed to distinguish epitopes from non-epitopes given the peptides (short chain of amino acids). Methods using classical machine learning require manual feature engineering on the primary sequence of peptides. Their mediocre performances indicate the challenge of B-cell epitope prediction~\cite{galanis2021linear}. Recently, several methods use embeddings derived from language models trained on large datasets of protein sequences to improve accuracy~\cite{bahai2021epitopevec,clifford2022bepipred,collatz2021epidope}. However, working with these huge models and neural embeddings is computationally expensive, especially for researchers with limited resources.

Antibodies can be classified into different types of immunoglobulins (Ig), each with different functions. Also, studies have shown that particular antigens induce specific types of antibodies~\cite{kuby}. For instance, IgA is vital against viral infections, IgE is involved in allergy, and IgM is linked to inflammation and autoimmunity. It is relevant to characterize epitopes potentially inducing specific classes of antibodies for applications like developing processing methods that mitigate food allergenicity. Only a few methods have been developed for Ig type-specific epitope classification, using classical machine learning~\cite{gupta2013identification,kadam2021antibody}.

In this work, we propose a new deep learning-based unified framework for the tasks of (non-)epitope prediction and Ig type-specific epitope classification. Unlike most existing tools that first compute sequence-level features for each peptide and then train a classifier, our end-to-end framework accepts variable-length sequences as input, encodes features at the residue level, and learns representations of peptides for classification. To our knowledge, no previous research has developed and trained Transformer-based networks for epitope prediction. We also incorporate cost-sensitive learning into our framework and design objective functions that handle the data imbalance, which is often overlooked in prior works. Experiments on data comprising over 120000 peptides obtained from the Immune Epitope Database (IEDB)~\cite{vita2019immune} show results exceeding state-of-the-art baselines in terms of prediction performance. Our framework achieves high predictive capacity with an area under the curve (AUC) of 86\% and outperforms the best baseline by 6\% in accuracy. Ablation studies demonstrate the usefulness of different components in the framework. More specifically, the main contributions of our work are summarized below.
\begin{itemize}
\item We propose a simple encoding method for amino acids, leveraging the eigen decomposition of an amino-acid scoring matrix.
\item We extend a logit adjustment technique and design a general loss function to address the class imbalance in binary and multiclass classifications.
\item A neural network based on Transformer is developed for peptide classification, without relying on large language models.
\item B-cell epitope data are collected and processed to create new redundancy-reduced datasets for benchmarking, concerning possible false negatives.
\end{itemize}

\section{Related Work}
\subsection{B-Cell Epitope Prediction}
Most of the machine learning-based methods designed to predict and classify B-cell epitopes are for linear epitopes rather than conformational epitopes, because of the more readily available data on protein primary sequence and by contrast the scarcity of data on protein three-dimensional structure. These methods vary from support vector machines (SVM)~\cite{bahai2021epitopevec,gupta2013identification,singh2013improved}, tree-based methods~\cite{jespersen2017bepipred,kadam2021antibody,manavalan2018ibce}, to neural networks~\cite{clifford2022bepipred,collatz2021epidope,liu2020deep,xu2022netbce}. No matter what kind of approaches they use, the key point is how to extract appropriate features from the epitope sequences as input for machine learning. The features used include the amino acid composition of the peptide~\cite{bahai2021epitopevec,gupta2013identification,kadam2021antibody,liu2020deep,manavalan2018ibce,singh2013improved} and propensity scales that depict the physicochemical properties of residues, including hydrophilicity, flexibility, surface accessibility, etc.~\cite{jespersen2017bepipred,manavalan2018ibce}. Some models have the limitation that they only process fixed-length sequences~\cite{gupta2013identification,liu2020deep,singh2013improved}. Besides, some do not address the similarity between sequences before splitting training and test sets~\cite{gupta2013identification,kadam2021antibody,liu2020deep}.

Given the analogy between amino acid sequences and human languages, natural language processing (NLP) techniques are applied in many biological property prediction tasks~\cite{ma2022identification,ofer2021language,teufel2022signalp,thumuluri2022deeploc}. With sufficient data, deep neural networks can automatically learn meaningful features, thus reducing the need for handcrafted features~\cite{lecun2015deep}. Recurrent neural networks (RNN), with long short-term memory (LSTM)~\cite{hochreiter1997long} as a representative, are dominant in NLP because their chain-like structures allow them to process over sequences without pre-specified constraints on the sequence lengths. In recent years, Transformer models have become the state of the art by using attention and eliminating the need for recurrent layers, thus overcoming the sequential bottleneck of RNNs~\cite{vaswani2017attention}. There is great interest in learning protein representation using language modeling, of which the paradigm is pre-training a large model in a self-supervised way on a large corpus of text, and then fine-tuning it in a supervised way for specific tasks~\cite{devlin2019bert}. Following the successful applications in protein property prediction, some studies use embeddings from protein language models as features to train classifiers for epitope prediction~\cite{bahai2021epitopevec,clifford2022bepipred,collatz2021epidope}, which require a large demand of computing resources and time. Moreover, pre-trained models for proteins may not be an optimal solution for peptides, which are typically much shorter than proteins.

\subsection{Imbalanced Learning}
In data mining, the imbalance problem occurs when the distribution of classes (labels) is not uniform. This poses a challenge for the prediction on minority classes and makes learning biased toward majority classes, especially when the distribution is highly skewed. Fundamental approaches to coping with imbalance can be broadly divided into re-sampling and re-weighting. Re-sampling modifies the datasets, for example by under-sampling or over-sampling~\cite{kang2019decoupling}. It is also used by existing epitope prediction methods~\cite{gupta2013identification,liu2020deep}. The drawbacks of re-sampling are that under-sampling incurs information loss in the majority classes, while over-sampling increases the training workload and can lead to overfitting for the minority classes. Alternatively, re-weighting modifies the model, for example by changing the loss function. The core idea is to adjust the weights of different samples in the loss, such as the misclassified ones and the under-represented ones~\cite{cui2019class,lin2017focal}. A recent paper proposed a strategy that modifies the inside of the logarithm in the standard cross-entropy loss and presented a statistical grounding for the strategy~\cite{menon2020long}.

\section{Methods}
\subsection{Task Definition and Solution Overview}
Given a linear peptide, which can be represented as a linear sequence of amino acids, our task contains two subtasks. The major one is to predict whether the peptide is an epitope or non-epitope and the minor one is to classify an epitope according to the specific class of Ig it potentially binds to. In addition, a score between 0 and 1 inclusive is given to indicate the probability of the peptide being an epitope. This allows users to choose different thresholds for determining epitopes, which is a common practice in epitope prediction.

\begin{figure}[htbp]
\centerline{\includegraphics[width=\textwidth]{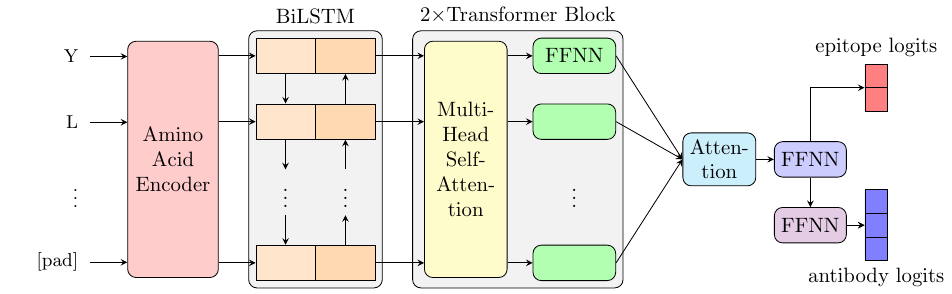}}
\caption{A general illustration of our framework. The input peptide has been tokenized. The network is trained by optimizing logit-adjusted losses.}
\label{fig1}
\end{figure}

The problem is framed as a binary classification for (non-)epitope prediction and a multiclass classification for Ig-specific epitope prediction. We preprocess B-cell epitope data from the IEDB for model training and evaluation. A rough overview of the proposed framework is shown in Fig.~\ref{fig1}. Raw peptide sequences are firstly tokenized at the residue level and then converted to numeric vectors by the encoder as input for the neural network. These numeric representations are passed through a bidirectional LSTM (BiLSTM) layer followed by two Transformer encoder blocks. Aggregated via an attention mechanism, the representations are classified using fully connected feedforward neural networks (FFNN). The whole model is jointly trained in a supervised manner for both epitope and Ig binding predictions, being aware of class imbalance.

\subsection{Tokenization and Encoding}
Like sentences in natural languages, a peptide can be represented as a sequence of characters, each representing a residue. Since machine learning models can only work directly with numbers, the raw sequences have to be transformed into numerical form before being fed to models. We follow the standard way in NLP to tokenize peptide sequences and convert tokens into integers indices. A vocabulary of size 20 is used, which contains tokens corresponding to the 20 kinds of standard amino acids. In addition, a [unk] token and a [pad] token are added to the vocabulary. A peptide is tokenized, i.e., split into residues, with uncertain amino acids in the raw sequence being replaced by [unk] token. For a batch of peptides of varying lengths, [pad] token is added at the end of shorter sequences to ensure they have the same length as the longest sequence in the batch so that models can process the input in batches.

Many protein prediction methods use one-hot encoding to encode the residues. The problem with this binary representation is that it does not reflect the biological similarities between different amino acids. We propose an alternative encoding method that exploits similarity scoring matrices based on observed alignments of proteins. Generally, an amino-acid scoring matrix is a symmetric matrix of size $20\times 20$, the entries of which are in the form of $\log\frac{q_{ij}}{p_ip_j}$, where $q_{ij}$ is the substitution frequency of two amino acids in a homologous sequence, and $p_i$ and $p_j$ are the background frequencies~\cite{pearson2013selecting}. The ratio provides a measure of the probability of two amino acids appearing in an alignment with a biological sense relative to appearing by chance, and therefore it captures the biological similarities between amino acids. We choose the widely used BLOSUM62 matrix and take the exponential of it, denoted by $B$. Note that $B$ is positive definite and eigen decomposition factorizes $B$ into the product of $U\Sigma U^\top$, where $U$ is orthonormal and $\Sigma$ is a diagonal matrix of eigenvalues (all positive). We propose the following encoding matrix $E$ in which each row represents an amino acid:
\begin{equation}
E=U\sqrt{\Sigma}.
\end{equation}
With this representation, the dot product of amino acid vectors corresponds to their biological similarity. We find this idea analogous to some works in NLP, in which word embedding methods maximize the dot product of similar words and implicitly factorize pointwise mutual information matrices~\cite{levy2014neural,mikolov2013distributed}. An advantage is that we can perform truncated decomposition by selecting only the top eigenvalues to get a lower-dimensional representation. We leverage the whole $\Sigma$ since the dimension is not high. An additional dimension is added for the [unk] token, which is represented by a one-hot vector orthogonal to other amino acid tokens. The [pad] token is represented by a zero vector. Consequently, a peptide of length $l$ is encoded as a matrix with size $l\times 21$. With this encoding, we intend to inject some biological heuristics to help the downstream learning.

\subsection{Neural Networks}
In this section, we depict the architecture of the neural network model, following a forward propagation through the model. The BiLSTM~\cite{graves2005framewise} layer combines two LSTM layers, one which processes the sequence in the forward direction, and one which processes the sequence in the backward direction. An LSTM layer can be regarded as multiple copies of the same LSTM cell, each passing information encoded in the hidden state from one step to the next. This chain-like structure is naturally applicable to sequential data. In particular, LSTM augments the hidden state with a memory cell state and gates that control this state. This leads to additive rather than multiplicative updates to the hidden states to alleviate the vanishing gradient problem in ordinary RNNs~\cite{hochreiter1997long}. For each residue, the hidden state vectors computed by the forward and backward layers are concatenated as the output of BiLSTM, hence taking into account the contextual information in the sequence. In our implementation, the [pad] tokens are not involved in computation and do not affect the output of other tokens.

Next, a Transformer encoder~\cite{vaswani2017attention} takes the output of BiLSTM as input. Since the output of BiLSTM already encodes the ordering information in the sequence, positional encoding in the original Transformer model is not needed. A Transformer encoder is a stack of multiple Transformer blocks, each of which is made up of multi-headed self-attention and FFNN. Self-attention~\cite{cheng2016long} enables the model to attend other relevant residues in the sequence when processing each residue, which may lead to improved representations of context. Multi-headed attention expands the model's capacity to capture different relationships. The outputs of the self-attention layer are fed to a position-wise FFNN layer, such that the same FFNN is separately applied to each residue. For both self-attention layer and FFNN layer, residual connection~\cite{he2016deep} is used, followed by layer normalization~\cite{ba2016layer}. The trainable parameters of each Transformer block are initialized according to Xavier uniform distribution~\cite{glorot2010understanding}. Let $d$ denote the size of an output vector of a residue. The output of the Transformer encoder has the same size of $l\times d$ as the input for a sequence of length $l$.

Attention mechanisms are often used in NLP to provide more flexibility in the context representation at the sentence or document level~\cite{yang2016hierarchical}. To acquire a sequence-level representation vector for the whole peptide based on the output of the previous module, we introduce an attention layer as a pooling layer to aggregate the information encoded in the residue vectors. Let vector $\mathbf{r_i}\in \mathbb{R}^{d}$ denotes the \textit{i}th residue in a peptide of length $l$, the peptide vector $\mathbf{p}$ is computed as a weighted sum of the residue vectors as follows:
\begin{align}
\alpha_i&=\frac{\exp(\mathbf{q}^\top\mathbf{r_i}/\sqrt{d})}{\sum_{j=1}^{l}\exp(\mathbf{q}^\top\mathbf{r_j}/\sqrt{d})}, \\
\mathbf{p}&=\sum_{i=1}^{l}\alpha_i\mathbf{r_i}.
\end{align}
That is, we use scaled dot-product attention~\cite{vaswani2017attention} and compute the attention weights using the softmax function. The query vector $\mathbf{q}$ is initialized such that its elements follow the standard normal distribution, scaled to avoid large variance in the products, and is jointly learned during the training. We implement the attention layer using masks to ignore padded positions in the sequence. The peptide vector $\mathbf{p}$ is used as features for epitope classification in the subsequent classifier. The classifier contains two heads of two-layer FFNNs for (non-)epitope and Ig-specific classification, respectively. Rectified linear unit (ReLU) activation~\cite{glorot2011deep} is used in the FFNNs.

\subsection{Loss Functions}
Our data may exhibit an imbalanced label distribution. It is desirable to learn a model that minimizes the balanced error, which averages each of the per-class errors, instead of the naïve misclassification error. We modify the standard cross-entropy losses, based on a logit adjustment technique~\cite{menon2020long} and focal loss~\cite{lin2017focal}. Both are originally proposed to address the imbalance problem in visual recognition.

One of the logit adjustment techniques is adding offsets to the logits in the loss function during training. We illustrate the intuition behind the logit-adjusted loss here. Suppose the unnormalized output (logits) of the model for all classes are $\mathbf{z}=[z_0,z_1,...,z_{C-1}]^\top$, where $C$ is the number of classes. Given a sample with instance $x$ and class label $y$ following distribution $\mathbb{P}$, the loss $\mathbb{E}_{(x,y)\sim\mathbb{P}}[\ell(\mathbf{z},y)]$ is minimized during training. The standard softmax cross-entropy for a sample $(x,y)$ is defined as $\ell(\mathbf{z}, y)=-\log(S(z_y))=-\log(\exp(z_y)/\sum_{i=0}^{C-1}\exp(z_i))$, where $S$ denotes the softmax function. One may view $S(z_y)\propto\exp(z_y)$ as an estimate of $\mathbb{P}(y|x)$, where $\mathbb{P}(y|x)\propto\mathbb{P}(x|y)\mathbb{P}(y)$ (Bayes' Theorem). However, to reduce the balanced error, balanced class-probability $\mathbb{P}_\mathtt{bal}(y|x)\propto\mathbb{P}(x|y)\cdot\frac{1}{C}$ instead of the standard $\mathbb{P}(y|x)$ should be used in Bayes-optimal prediction. Noticing $\mathbb{P}_\mathtt{bal}(y|x)\propto\mathbb{P}(y|x)/\mathbb{P}(y)$, the following logit-adjusted softmax cross-entropy loss was proposed in~\cite{menon2020long}:
\begin{equation}\label{eq4}
\ell_\mathtt{softmax}(\mathbf{z},y)=-\log\frac{\exp(z_y+\tau\log\pi_y)}{\sum_{i=0}^{C-1}\exp(z_i+\tau\log\pi_i)},
\end{equation}
where $\pi_y$ are empirical class frequencies used to estimate priors $\mathbb{P}(y)$, and $\tau\ge 0$ is a scaling parameter. In this way, the model directly estimates $\mathbb{P}_\mathtt{bal}(y|x)$ using $z_y$, while it can still be trained with cross-entropy loss. Note that the prediction is still $\arg\max_y z_y$ as usual.

The original logit adjustment is for softmax loss in multiclass classification. We extend it to modify sigmoid loss and generalize it for binary and multi-label classification. In binary classification, suppose the logit for sample $(x,y)$ is $z$. Sigmoid is equivalent to softmax when $C=2$ and $\mathbf{z}=[0, z]^\top$, in that $\sigma(z)=1/(1+e^{-z})=e^z/(e^z+e^0)=S(z)$ and $1-\sigma(z)=S(0)$, where $\sigma$ denotes the sigmoid function. Given this connection, following the idea on softmax loss, we can derive the logit-adjusted sigmoid cross-entropy loss for binary classification:
\begin{equation}
\ell_\mathtt{binary}(z,y)=-y\log(\sigma(z+\tau\log\frac{\pi}{1-\pi}))-(1-y)\log(1-\sigma(z+\tau\log\frac{\pi}{1-\pi})),
\end{equation}
where $\pi$ is the empirical positive class frequency. Here we add the logarithm of odds to the logit, in contrast to adding the logarithm of probability in logit-adjusted softmax. In the balanced scenario ($\pi=0.5$), it becomes the standard sigmoid loss.

We can treat multiclass classification as multiple one-vs-all binary classification tasks, i.e. multi-label classification, and thus use sigmoid loss. Unlike softmax loss, sigmoid loss does not assume mutual exclusiveness among each class. This aligns well with real-world data, where different classes might have some overlaps. Using the same notation as Equation \eqref{eq4}, for convenience, we define adjusted logits $z_i^\ast$ as:
\begin{equation}
z_i^\ast=
\begin{cases}
z_i+\tau\log\frac{\pi_i}{1-\pi_i} &\text{if}\ i=y \\
-z_i-\tau\log\frac{\pi_i}{1-\pi_i} &\text{otherwise}.
\end{cases}
\end{equation}
The logit-adjusted sigmoid cross-entropy loss for multiclass classification is:
\begin{equation}\label{eq7}
\ell_\mathtt{sigmoid}(\mathbf{z},y)=-\frac{1}{C}\sum_{i=0}^{C-1}\log\frac{1}{1+\exp(-z_i^\ast)}.
\end{equation}
Another approach is training with standard sigmoid loss and then adjusting the logits to predict. The scaling parameter may be tuned in a post-hoc way, without training with different $\tau$. Similar to the procedure in~\cite{menon2020long}, we instead predict:
\begin{equation}\label{eq8}
\arg\max_y z_y-\tau\log\frac{\pi_y}{1-\pi_y}.
\end{equation}

Furthermore, we apply focal loss~\cite{lin2017focal} to Equation \eqref{eq4} and \eqref{eq7}. Denote $p_y=S(z_y+\tau\log\pi_y)$, the focal softmax loss can be written as:
\begin{equation}
\ell_\mathtt{focal-softmax}(\mathbf{z},y)=-(1-p_y)^\gamma\log p_y,
\end{equation}
and denote $p_i^\ast=\sigma(z_i^\ast)$, the focal sigmoid loss can be written as:
\begin{equation}\label{eq10}
\ell_\mathtt{focal-sigmoid}(\mathbf{z},y)=-\frac{1}{C}\sum_{i=0}^{C-1}(1-p_i^\ast)^\gamma\log p_i^\ast,
\end{equation}
where $\gamma\ge 0$ is a focusing parameter. Intuitively, since the modulating factor $(1-p)^\gamma$ becomes smaller when $p$ is closer to 1, this focal term reduces the relative loss for well-classified samples, putting emphasis on the difficult samples. The losses defined above are quite flexible. When $\gamma=0$ and $\tau=0$, they are equivalent to the standard cross-entropy. Typically, we set $\gamma=1$ and $\tau=1$.

We adopt the proposed loss functions in our task. The total cost function is defined as $L=\alpha_\mathtt{p}L_\mathtt{p}+\alpha_\mathtt{ig}L_\mathtt{ig}$, where $L_\mathtt{p}$ is (non-)epitope classification loss and $L_\mathtt{ig}$ is Ig-specific classification loss, averaging over all training samples. The coefficient can be set according to the relative importance of the subtasks in practice, e.g., setting $\alpha_\mathtt{ig} = 0$ for only training epitope prediction. We use AdamW~\cite{loshchilov2017decoupled} with AMSGrad~\cite{reddi2018convergence} as the optimization algorithm to minimize the total loss $L$. The learning rate is scheduled such that during training it increases linearly from zero to a specified value in the warmup period~\cite{goyal2017accurate}, followed by a cosine decay~\cite{loshchilov2016sgdr}. The purpose of this dynamic learning rate is to reduce the instability at the early stage of optimization, avoid oscillation and help the model converge to a local minimum near the end of optimization. Regularization techniques such as dropout~\cite{srivastava2014dropout} and weight decay are applied to prevent overfitting.

\section{Data and Experiments}
\subsection{Datasets and Preprocessing}
The IEDB catalogs experimental data on B-cell and T-cell epitopes studied in humans and other species in the context of infectious disease, allergy, autoimmunity, etc., curated from the scientific literature~\cite{vita2019immune}. To our knowledge, it is the most comprehensive epitope database containing the largest number of experimentally verified (non-)epitopes. We downloaded all B-cell epitope data for cleansing, which contains over 1.3 million entries of B-cell assays, associated with around 0.6 million epitopes. Note that there is a many-to-one relationship between assay and epitope. The data were processed in several steps as below.

We extracted linear peptides whose sequence contains only one-letter symbols, discarding peptides that contain modified residues. Peptides of length not larger than 25 were selected. The upper limit was set to reduce noise caused by the curation into IEDB, as long peptides could be epitope containing regions instead of exact epitopes. We counted the number of assays with positive and negative outcomes for each peptide. Following the instruction of IEDB, peptides having at least one positive measurement are defined as epitopes, and peptides having only negative measurements are defined as non-epitopes. Note that an epitope can have some negative assays. We further grouped epitopes by the type of antibodies they bind to in positive assays. Among the five major types of antibodies, there is no data on IgD. IgG is the predominant type that most epitopes induce. A few epitopes induce more than one type among IgA, IgE, and IgM. We labeled epitopes that specifically induce one of these three types but not the other two, while others were not used in the Ig prediction subtask.

Taking into account the homology between sequences, we utilized CD-HIT~\cite{fu2012cd}, a tool that uses a greedy incremental clustering algorithm and outputs the longest representative sequence for each cluster. We clustered epitopes and non-epitopes respectively using an identity threshold of 0.8 and removed redundant sequences. The sequences in non-epitopes that are similar to epitopes were further removed using the same threshold. There are several benefits of reducing redundancy. First, it ensures training and test sets do not have near identical sequences after splitting. Second, it reduces the bias of overrepresented sequences in training. Third, many short peptides are removed, especially in non-epitopes. It is beneficial since research shows that short peptides give false negative results in experiments, which confound computational epitope prediction~\cite{rahman2016inadequate}.

We filtered the peptides of the organism severe acute respiratory syndrome coronavirus 2 (SARS-CoV-2) to create a COVID dataset for a case study of our framework. In the rest of the data, there are 64019 non-epitopes and 64940 epitopes, among which there are 443 IgA epitopes, 1450 IgE epitopes, and 7715 IgM epitopes. 5000 non-epitopes were randomly sampled, and 5000 epitopes were stratified sampled according to the Ig label frequencies, resulting in a hold-out test set of size 10000. We performed the same sampling to create a validation set of size 10000. The remaining data constitute the training set.

\subsection{Baselines and Ablation Studies}
For the (non-)epitope prediction subtask, we choose several recently published machine learning-based methods as baselines, which not only have sufficient implementation details but also state improvement over major methods. We also use the publicly available dataset of NetBCE~\cite{xu2022netbce} for comparison. The dataset was compiled from the IEDB but curated with different criteria from ours to select epitopes and reduce homology, resulting in 97784 non-epitopes and 27095 epitopes, having a ratio of 3.6 between the numbers. Other public datasets are not used since they are either not reduced or contain much fewer peptides.

DLBEpitope~\cite{liu2020deep} uses dipeptide composition as the feature vector for peptides. Dipeptide composition is represented by a vector of length 400, specifying the fractions of all possible combinations of amino acid pairs in a peptide. The classifier is an FFNN with four hidden layers. RMSprop algorithm~\cite{graves2013generating} is used to optimize the cross-entropy.

EpiDope~\cite{collatz2021epidope} combines neural networks and a protein language model. Besides a widely used module composed of an embedding, a BiLSTM, and a linear layer, the architecture also involves a pre-trained model. The outputs of these two modules are concatenated and fed to an FFNN for classification. The model is trained as a whole with the weights of the language module being frozen.

NetBCE~\cite{xu2022netbce} applies one-hot encoding on the residues and uses a neural network to extract representations and classify peptides, with cross-entropy optimized by Adam~\cite{kingma2014adam} algorithm. The model contains a convolutional neural network (CNN) to capture the pattern in the sequences, followed by a BiLSTM layer to catch long-range dependencies. The CNN module is composed of a convolution~\cite{lecun1989backpropagation}, a batch normalization~\cite{ioffe2015batch}, and a max pooling layer.

Classical machine learning-based methods are also compared. LBtope~\cite{singh2013improved} is the first method that uses validated non-epitopes in training. It uses SVM on composition-based features. iBCE-EL~\cite{manavalan2018ibce} is a framework that contains several classifiers on the amino acid compositions and the physicochemical properties. For a fair comparison, we select their best model, extremely randomized tree, and use within ensemble learning, a variant of random forest, as a baseline.

For the Ig-specific epitope classification subtask, we choose IgPred~\cite{gupta2013identification} and AbCPE~\cite{kadam2021antibody} as the baselines. They are used to predict the antibody class based on the input epitope sequences that are experimentally verified. IgPred trains SVM using the radial basis function kernel while AbCPE trains AdaBoost classifier on the dipeptide composition features of the epitope sequences. We improve the performance by replacing AdaBoost with XGBoost~\cite{chen2016xgboost} and use it instead for comparison. Class weights are computed to account for the imbalance.

We conduct ablation studies to investigate the contributions of different parts to the overall performance of our framework. We experiment with different sequence models. Particularly, besides the hybrid model, we also try a two-layer BiLSTM and a four-layer Transformer encoder with positional encoding being applied to the input. Also, our proposed encoding method and loss functions are compared with conventional methods. All the models in this part are trained for 100 epochs, using the AdamW optimizer with a warmup period of 200 steps.

In epitope prediction, the AUC of receiver operating characteristic (ROC), summarizing the tradeoff between sensitivity and specificity, is used in nearly all the papers for performance evaluation. Therefore we select AUC as the main metric for the (non-)epitope prediction subtask. For the imbalanced Ig-specific prediction subtask, balanced accuracy (Acc), i.e. the average of recall in each class, is used as the metric. We try to attain a comparable AUC for epitope prediction when comparing different loss functions.

We perform hyperparameter tuning and monitor the performance on the validation set during training to avoid overfitting and underfitting. Hyperparameters include learning rate, batch size, weight decay coefficient, dropout probability, hidden dimension size, number of heads in Transformer, etc. We evaluate the models on the test set. The results are averaged across five runs with different initialization of model parameters using different random seeds. All models are implemented in PyTorch~\cite{paszke2019pytorch} and scikit-learn~\cite{pedregosa2011scikit}. Experiments are done in Ubuntu 20.04 with Intel Xeon 2.20 GHz CPU or NVIDIA Tesla P100 GPU.

\section{Results and Analysis}
Table~\ref{tab1} shows that our model achieves better performance than baselines. Traditional models require manually designed and computed features, yet still do not perform better than other sequence-based models. The primary structure, i.e. the linear sequence of amino acids, of a peptide greatly determines the high-level structures and functions of the peptide. It is analogous to how the arrangements of words define the semantic meaning of a sentence in a context-dependent way. Therefore, it is reasonable that employing NLP techniques can help understand the information encoded in peptides.
Unlike NetBCE, we do not use CNN in our models. The rationale is that compared to self-attention in Transformer, convolution has a limited receptive field, which depends on the kernel size, and does not flexibly adapt to the input content due to the static weights of the filter.
\begin{table}[htbp]
\caption{Performance (\%) of baselines and our model for epitope prediction and Ig-specific classification. Models are ordered by publication dates.}
\begin{center}
\begin{tabular}{l|l|l|l|l|l}
\hline
\multirow[t]{2}{*}{Model} & \multicolumn{3}{l|}{Our dataset} & \multicolumn{2}{l}{NetBCE dataset} \\
\cline{2-6}
 & AUC & Acc & Acc (Ig) & AUC & Acc \\
\hline
LBtope~\cite{singh2013improved} & 78.84 & 71.80 & - & 81.70 & 61.93 \\
IgPred~\cite{gupta2013identification} & - & - & 70.14 & - & - \\
iBCE-EL~\cite{manavalan2018ibce} & 80.38 & 73.05 & - & 83.17 & 64.76 \\
DLBEpitope~\cite{liu2020deep} & 76.39 & 69.04 & - & 82.94 & 69.92 \\
EpiDope~\cite{collatz2021epidope} & 81.62 & 73.19 & - & 83.44 & 70.72 \\
AbCPE~\cite{kadam2021antibody} & - & - & 70.27 & - & - \\
NetBCE~\cite{xu2022netbce} & 82.41 & 72.80 & - & 84.00 & 68.65 \\
Ours & \textbf{85.80} & \textbf{77.29} & \textbf{72.21} & \textbf{86.10} & \textbf{74.81} \\
\hline
\end{tabular}
\label{tab1}
\end{center}
\end{table}

Table~\ref{tab2} shows the results of ablation studies regarding our proposed model and its variants on the test set. Our simple encoding method for amino acids yields an improvement in performance for all three architectures, compared to the conventional method that uses an embedding layer with learnable weights. Incorporating this additional biological information is helpful in downstream learning.
\newcommand{\xmark}{\scalebox{0.75}{\usym{2613}}}
\begin{table}[htbp]
\caption{Performance (\%) of different models and encodings in ablation studies.}
\begin{center}
\begin{tabular}{l|c|l|l|l}
\hline
Architecture & \multirow{2}{1.3cm}{Residue encoding} & AUC & Acc & Acc (Ig) \\
& & & & \\
\hline
BiLSTM$\times$2 & \xmark & 84.04 & 75.58 & 67.00 \\
BiLSTM$\times$2 & \checkmark & 85.06 & 76.23 & 69.86 \\
Transformer$\times$4 & \xmark & 84.99 & 76.57 & 66.98 \\
Transformer$\times$4 & \checkmark & 85.76 & 77.17 & 68.06 \\
BiLSTM+Transformer$\times$2 & \xmark & 85.27 & 76.73 & 70.73 \\
BiLSTM+Transformer$\times$2 & \checkmark & 85.80 & 77.29 & 72.21 \\
\hline
\end{tabular}
\label{tab2}
\end{center}
\end{table}
Interestingly, the hybrid architecture performs better than pure LSTM and pure Transformer encoder. Typically, Transformer needs a lot of data to overcome its relative lack of inductive bias. This could explain why a pure Transformer encoder does not perform well on Ig-specific epitope classification. For a fair comparison, all of our sequence-based models are constructed such that they cost roughly the same training time. Generally, the time complexity for a recurrent layer is $O(ld^2)$, while the time complexity for a Transformer block is $O(l^2d)$, where $l$ is the sequence length and $d$ is the hidden dimension size~\cite{vaswani2017attention}. Fortunately, $l$ is small in our task, with an average of approximately 15, and we use $d=64$ in LSTM layers and $d=128$ in Transformer blocks.

\begin{figure}[htbp]
\centerline{\includegraphics[width=0.5\textwidth]{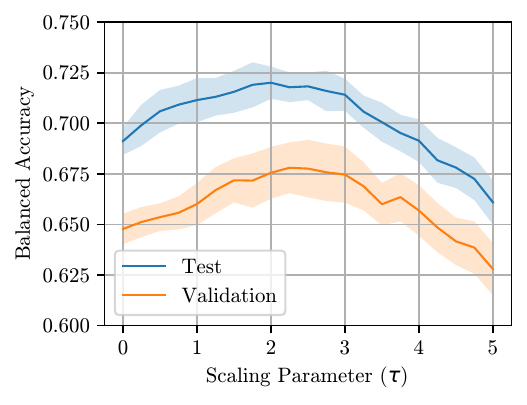}}
\caption{Post-hoc adjustment with varying scaling parameter on the logits of Ig-specific classification. The models are trained using standard sigmoid loss.}
\label{fig2}
\end{figure}

Table~\ref{tab3} shows the performance of our best model on the test set with different loss functions in Ig-specific epitope classification, which has an imbalance ratio of approximately 17. The standard cross-entropy losses have a strong bias towards the majority class IgM. Logit adjustment technique is conducive for both standard sigmoid and softmax loss. Focal loss further balances the per-class accuracy, though it does not affect the balanced accuracy very much.
\begin{table}[htbp]
\caption{Accuracies (\%) of Ig-specific classification with different loss functions.}
\begin{center}
\begin{tabular}{l|l|l|l|l}
\hline
Loss function & Balanced & IgA & IgE & IgM \\
\hline
softmax & 68.74 & 64.71 & 47.99 & 93.52 \\
logit-adjusted softmax & 70.78 & 70.59 & 51.79 & 89.98 \\
focal logit-adjusted softmax & 71.20 & 70.59 & 53.12 & 89.90 \\
sigmoid & 68.79 & 60.78 & 52.08 & 93.49 \\
logit-adjusted sigmoid & 72.01 & 68.63 & 56.25 & 91.13 \\
focal logit-adjusted sigmoid & 72.21 & 74.12 & 55.00 & 87.51 \\
\hline
\end{tabular}
\label{tab3}
\end{center}
\end{table}
Overall, our proposed logit-adjusted sigmoid loss (Equation \eqref{eq7} and \eqref{eq10}) achieves better performance than logit-adjusted softmax loss. Also, it is versatile in that it is applicable to binary ($C=1$), multiclass, and multi-label classification. Alternatively, we can apply the post-hoc approach and predict as per \eqref{eq8}. Fig.~\ref{fig2} shows the effect of tuning the parameter. With suitable tuning, the accuracy is on par with using the logit-adjusted loss.

\section{Application and Discussion}
\begin{figure}[htbp]
\centerline{\includegraphics[width=0.5\textwidth]{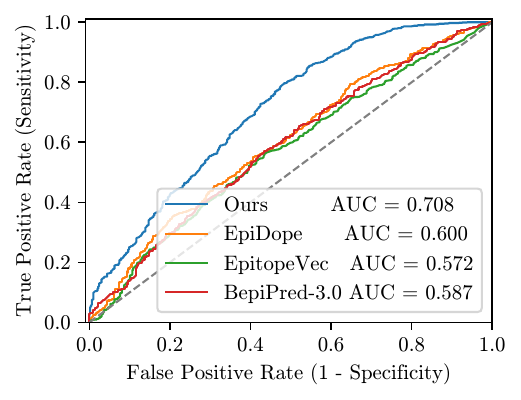}}
\caption{ROC curves of different tools for epitope prediction on the COVID dataset.}
\label{fig3}
\end{figure}

We further demonstrate the validity of our framework in application on the small hold-out COVID dataset, which contains 497 non-epitopes and 1180 epitopes, and show the ROC curves in Fig.~\ref{fig3}. Three recently published frameworks called EpiDope~\cite{collatz2021epidope}, EpitopeVec~\cite{bahai2021epitopevec}, and BepiPred-3.0~\cite{clifford2022bepipred} are used for comparison, all of which use embeddings produced by language models, with AUC ranging from 63\% to 78\% reported in their papers. We directly input the COVID dataset to these tools. For the tools outputting scores per residue, we average to obtain a score for the peptide and compute the AUC. A decrease in performance is observed for all four frameworks compared to the reported AUC on their test sets. A possible explanation is that epitopes from different organisms could have different underlying data distributions, making a general model trained on a variety of species underperform on specific organisms~\cite{ashford2021organism}. Nevertheless, our framework still significantly outperforms the other three. Without a high computational cost incurred by the large models, the inference time is only 5 seconds on this dataset, while other tools using language models typically take over minutes.

The experimental results show that our framework is promising as a pre-screening tool for prioritizing targets for laboratory investigation. Our lightweight model can be trained on hundreds of thousands of sequences in moderate time, spending a couple of hours on CPU. Thus, it is friendly to researchers having limited resources such as only CPU or low-end GPU. Moreover, based on the effectiveness and flexibility of the proposed methods, we believe our framework has the potential to be applied to other peptide classification problems. With the increasing availability of sequence and structure data, the information of antibodies can be incorporated in the future to model antigen-antibody interactions. Such studies will provide insights into the ligand-receptor interactions during immune response~\cite{li2021structure} and benefit the research on individualized immunotherapy~\cite{widrich2020modern}.

\section{Conclusion}
This paper presents a deep learning-based multi-task framework, which is called BeeTLe, for linear B-cell epitope prediction and antibody type-specific epitope classification using Transformer and LSTM encoders. It involves a simple yet effective residue encoding method, a model whose backbone combines recurrent layers and attention mechanisms to learn feature representations for peptides, and modified cross-entropy loss functions to address the imbalance problem. A large dataset with potential false-negative epitopes being reduced is curated for benchmarking. We implement and deploy a command-line tool to facilitate the use and extension of our work. The code and the data are open-source at \url{https://github.com/yuanx749/bcell}.

\subsubsection{Acknowledgements.}
The author sincerely thanks all the reviewers for their constructive feedback, and Jiarong Liang for the valuable discussions on the concepts and tools in immunology.

%
%
%
\bibliographystyle{splncs04}
\bibliography{bcell}

@article{hochreiter1997long,
  doi={10.1162/neco.1997.9.8.1735},
  title={Long short-term memory},
  author={Hochreiter, Sepp and Schmidhuber, J{\"u}rgen},
  journal={Neural computation},
  volume={9},
  number={8},
  pages={1735--1780},
  year={1997},
  publisher={MIT Press}
}

@inproceedings{vaswani2017attention,
  title={Attention is all you need},
  author={Vaswani, Ashish and Shazeer, Noam and Parmar, Niki and Uszkoreit, Jakob and Jones, Llion and Gomez, Aidan N and Kaiser, {\L}ukasz and Polosukhin, Illia},
  booktitle={Proceedings of the 31st International Conference on Neural Information Processing Systems},
  pages={6000--6010},
  year={2017}
}

@inproceedings{devlin2019bert,
  doi={10.18653/v1/N19-1423},
  title={BERT: Pre-training of Deep Bidirectional Transformers for Language Understanding},
  author={Devlin, Jacob and Chang, Ming-Wei and Lee, Kenton and Toutanova, Kristina},
  booktitle={Proceedings of the 2019 Conference of the North American Chapter of the Association for Computational Linguistics: Human Language Technologies, Volume 1 (Long and Short Papers)},
  pages={4171--4186},
  year={2019}
}

@inproceedings{kang2019decoupling,
  title={Decoupling Representation and Classifier for Long-Tailed Recognition},
  author={Kang, Bingyi and Xie, Saining and Rohrbach, Marcus and Yan, Zhicheng and Gordo, Albert and Feng, Jiashi and Kalantidis, Yannis},
  booktitle={International Conference on Learning Representations},
  year={2020}
}

@inproceedings{lin2017focal,
  doi={10.1109/ICCV.2017.324},
  title={Focal Loss for Dense Object Detection},
  author={Lin, Tsung-Yi and Goyal, Priya and Girshick, Ross and He, Kaiming and Doll{\'a}r, Piotr},
  booktitle={2017 IEEE International Conference on Computer Vision (ICCV)},
  pages={2999--3007},
  year={2017},
  organization={IEEE}
}

@inproceedings{cui2019class,
  doi={10.1109/CVPR.2019.00949},
  title={Class-Balanced Loss Based on Effective Number of Samples},
  author={Cui, Yin and Jia, Menglin and Lin, Tsung-Yi and Song, Yang and Belongie, Serge},
  booktitle={2019 IEEE/CVF Conference on Computer Vision and Pattern Recognition (CVPR)},
  pages={9260--9269},
  year={2019},
  organization={IEEE}
}

@inproceedings{menon2020long,
  title={Long-tail learning via logit adjustment},
  author={Menon, Aditya Krishna and Jayasumana, Sadeep and Rawat, Ankit Singh and Jain, Himanshu and Veit, Andreas and Kumar, Sanjiv},
  booktitle={International Conference on Learning Representations},
  year={2021}
}

@inproceedings{mikolov2013distributed,
  title={Distributed representations of words and phrases and their compositionality},
  author={Mikolov, Tomas and Sutskever, Ilya and Chen, Kai and Corrado, Greg and Dean, Jeffrey},
  booktitle={Proceedings of the 26th International Conference on Neural Information Processing Systems-Volume 2},
  pages={3111--3119},
  year={2013}
}

@inproceedings{levy2014neural,
  title={Neural word embedding as implicit matrix factorization},
  author={Levy, Omer and Goldberg, Yoav},
  booktitle={Proceedings of the 27th International Conference on Neural Information Processing Systems-Volume 2},
  pages={2177--2185},
  year={2014}
}

@article{graves2005framewise,
  doi={10.1016/j.neunet.2005.06.042},
  title={Framewise phoneme classification with bidirectional LSTM and other neural network architectures},
  author={Graves, Alex and Schmidhuber, J{\"u}rgen},
  journal={Neural networks},
  volume={18},
  number={5-6},
  pages={602--610},
  year={2005},
  publisher={Elsevier}
}

@inproceedings{cheng2016long,
  doi={10.18653/v1/D16-1053},
  title={Long Short-Term Memory-Networks for Machine Reading},
  author={Cheng, Jianpeng and Dong, Li and Lapata, Mirella},
  booktitle={Proceedings of the 2016 Conference on Empirical Methods in Natural Language Processing},
  pages={551--561},
  year={2016}
}

@inproceedings{he2016deep,
  doi={10.1109/CVPR.2016.90},
  title={Deep Residual Learning for Image Recognition},
  author={He, Kaiming and Zhang, Xiangyu and Ren, Shaoqing and Sun, Jian},
  booktitle={2016 IEEE Conference on Computer Vision and Pattern Recognition (CVPR)},
  pages={770--778},
  year={2016},
  organization={IEEE}
}

@article{ba2016layer,
  title={Layer normalization},
  author={Ba, Jimmy Lei and Kiros, Jamie Ryan and Hinton, Geoffrey E},
  journal={arXiv preprint arXiv:1607.06450},
  year={2016}
}

@inproceedings{glorot2010understanding,
  title={Understanding the difficulty of training deep feedforward neural networks},
  author={Glorot, Xavier and Bengio, Yoshua},
  booktitle={Proceedings of the thirteenth international conference on artificial intelligence and statistics},
  pages={249--256},
  year={2010},
  organization={JMLR Workshop and Conference Proceedings}
}

@inproceedings{yang2016hierarchical,
  doi={10.18653/v1/N16-1174},
  title={Hierarchical attention networks for document classification},
  author={Yang, Zichao and Yang, Diyi and Dyer, Chris and He, Xiaodong and Smola, Alex and Hovy, Eduard},
  booktitle={Proceedings of the 2016 conference of the North American chapter of the association for computational linguistics: human language technologies},
  pages={1480--1489},
  year={2016}
}

@inproceedings{glorot2011deep,
  title={Deep sparse rectifier neural networks},
  author={Glorot, Xavier and Bordes, Antoine and Bengio, Yoshua},
  booktitle={Proceedings of the fourteenth international conference on artificial intelligence and statistics},
  pages={315--323},
  year={2011},
  organization={JMLR Workshop and Conference Proceedings}
}

@inproceedings{loshchilov2017decoupled,
  title={Decoupled Weight Decay Regularization},
  author={Loshchilov, Ilya and Hutter, Frank},
  booktitle={International Conference on Learning Representations},
  year={2019}
}

@inproceedings{reddi2018convergence,
  title={On the Convergence of Adam and Beyond},
  author={Reddi, Sashank J and Kale, Satyen and Kumar, Sanjiv},
  booktitle={International Conference on Learning Representations},
  year={2018}
}

@article{goyal2017accurate,
  title={Accurate, large minibatch sgd: Training imagenet in 1 hour},
  author={Goyal, Priya and Doll{\'a}r, Piotr and Girshick, Ross and Noordhuis, Pieter and Wesolowski, Lukasz and Kyrola, Aapo and Tulloch, Andrew and Jia, Yangqing and He, Kaiming},
  journal={arXiv preprint arXiv:1706.02677},
  year={2017}
}

@inproceedings{loshchilov2016sgdr,
  title={SGDR: Stochastic Gradient Descent with Warm Restarts},
  author={Loshchilov, Ilya and Hutter, Frank},
  booktitle={International Conference on Learning Representations},
  year={2017}
}

@article{srivastava2014dropout,
  title={Dropout: a simple way to prevent neural networks from overfitting},
  author={Srivastava, Nitish and Hinton, Geoffrey and Krizhevsky, Alex and Sutskever, Ilya and Salakhutdinov, Ruslan},
  journal={The Journal of Machine Learning Research},
  volume={15},
  number={1},
  pages={1929--1958},
  year={2014},
  publisher={JMLR. org}
}

@article{graves2013generating,
  title={Generating sequences with recurrent neural networks},
  author={Graves, Alex},
  journal={arXiv preprint arXiv:1308.0850},
  year={2013}
}

@inproceedings{kingma2014adam,
  title={Adam: A method for stochastic optimization},
  author={Kingma, Diederik P and Ba, Jimmy},
  booktitle={International Conference on Learning Representations},
  year={2015}
}

@article{lecun1989backpropagation,
  doi={10.1162/neco.1989.1.4.541},
  title={Backpropagation applied to handwritten zip code recognition},
  author={LeCun, Yann and Boser, Bernhard and Denker, John S and Henderson, Donnie and Howard, Richard E and Hubbard, Wayne and Jackel, Lawrence D},
  journal={Neural computation},
  volume={1},
  number={4},
  pages={541--551},
  year={1989},
  publisher={MIT Press}
}

@inproceedings{ioffe2015batch,
  title={Batch normalization: Accelerating deep network training by reducing internal covariate shift},
  author={Ioffe, Sergey and Szegedy, Christian},
  booktitle={International conference on machine learning},
  pages={448--456},
  year={2015},
  organization={pmlr}
}

@inproceedings{chen2016xgboost,
  doi={10.1145/2939672.2939785},
  title={Xgboost: A scalable tree boosting system},
  author={Chen, Tianqi and Guestrin, Carlos},
  booktitle={Proceedings of the 22nd acm sigkdd international conference on knowledge discovery and data mining},
  pages={785--794},
  year={2016}
}

@inproceedings{paszke2019pytorch,
  title={PyTorch: an imperative style, high-performance deep learning library},
  author={Paszke, Adam and Gross, Sam and Massa, Francisco and Lerer, Adam and Bradbury, James and Chanan, Gregory and Killeen, Trevor and Lin, Zeming and Gimelshein, Natalia and Antiga, Luca and others},
  booktitle={Proceedings of the 33rd International Conference on Neural Information Processing Systems},
  pages={8026--8037},
  year={2019}
}

@article{pedregosa2011scikit,
  title={Scikit-learn: Machine Learning in Python},
  author={Pedregosa, Fabian and Varoquaux, Ga{\"e}l and Gramfort, Alexandre and Michel, Vincent and Thirion, Bertrand and Grisel, Olivier and Blondel, Mathieu and Prettenhofer, Peter and Weiss, Ron and Dubourg, Vincent and others},
  journal={The Journal of Machine Learning Research},
  volume={12},
  pages={2825--2830},
  year={2011},
  publisher={JMLR. org}
}

@inproceedings{li2021structure,
  doi={10.1145/3447548.3467311},
  title={Structure-aware interactive graph neural networks for the prediction of protein-ligand binding affinity},
  author={Li, Shuangli and Zhou, Jingbo and Xu, Tong and Huang, Liang and Wang, Fan and Xiong, Haoyi and Huang, Weili and Dou, Dejing and Xiong, Hui},
  booktitle={Proceedings of the 27th ACM SIGKDD Conference on Knowledge Discovery \& Data Mining},
  pages={975--985},
  year={2021}
}

@inproceedings{widrich2020modern,
  title={Modern hopfield networks and attention for immune repertoire classification},
  author={Widrich, Michael and Sch{\"a}fl, Bernhard and Pavlovi{\'c}, Milena and Ramsauer, Hubert and Gruber, Lukas and Holzleitner, Markus and Brandstetter, Johannes and Sandve, Geir Kjetil and Greiff, Victor and Hochreiter, Sepp and others},
  booktitle={Proceedings of the 34th International Conference on Neural Information Processing Systems},
  pages={18832--18845},
  year={2020}
}

@article{potocnakova2016introduction,
  doi={10.1155/2016/6760830},
  title={An introduction to B-cell epitope mapping and in silico epitope prediction},
  author={Potocnakova, Lenka and Bhide, Mangesh and Pulzova, Lucia Borszekova},
  journal={Journal of immunology research},
  volume={2016},
  year={2016},
  publisher={Hindawi}
}

@article{sanchez2017fundamentals,
  doi={10.1155/2017/2680160},
  title={Fundamentals and methods for T-and B-cell epitope prediction},
  author={Sanchez-Trincado, Jose L and Gomez-Perosanz, Marta and Reche, Pedro A and others},
  journal={Journal of immunology research},
  volume={2017},
  year={2017},
  publisher={Hindawi}
}

@article{galanis2021linear,
  doi={10.3390/ijms22063210},
  title={Linear B-cell epitope prediction for in silico vaccine design: A performance review of methods available via command-line interface},
  author={Galanis, Kosmas A and Nastou, Katerina C and Papandreou, Nikos C and Petichakis, Georgios N and Pigis, Diomidis G and Iconomidou, Vassiliki A},
  journal={International journal of molecular sciences},
  volume={22},
  number={6},
  pages={3210},
  year={2021},
  publisher={MDPI}
}

@article{collatz2021epidope,
  doi={10.1093/bioinformatics/btaa773},
  title={EpiDope: a deep neural network for linear B-cell epitope prediction},
  author={Collatz, Maximilian and Mock, Florian and Barth, Emanuel and H{\"o}lzer, Martin and Sachse, Konrad and Marz, Manja},
  journal={Bioinformatics},
  volume={37},
  number={4},
  pages={448--455},
  year={2021},
  publisher={Oxford University Press}
}

@article{bahai2021epitopevec,
  doi={10.1093/bioinformatics/btab467},
  title={EpitopeVec: linear epitope prediction using deep protein sequence embeddings},
  author={Bahai, Akash and Asgari, Ehsaneddin and Mofrad, Mohammad RK and Kloetgen, Andreas and McHardy, Alice C},
  journal={Bioinformatics},
  volume={37},
  number={23},
  pages={4517--4525},
  year={2021},
  publisher={Oxford University Press}
}

@article{clifford2022bepipred,
  doi={10.1002/pro.4497},
  title={BepiPred-3.0: Improved B-cell epitope prediction using protein language models},
  author={Clifford, Joakim and H{\o}ie, Magnus Haraldson and Deleuran, Sebastian and Peters, Bjoern and Nielsen, Morten and Marcatili, Paolo},
  journal={Protein Science},
  pages={e4497},
  year={2022},
  publisher={Wiley Online Library}
}

@book{kuby,
  author = {Punt, Jenni},
  edition = {Eighth},
  isbn = {9781319114701},
  publisher = {W. H. Freeman and Company},
  title = {Kuby immunology},
  year = {2019},
}

@article{gupta2013identification,
  doi={10.1186/1745-6150-8-27},
  title={Identification of B-cell epitopes in an antigen for inducing specific class of antibodies},
  author={Gupta, Sudheer and Ansari, Hifzur Rahman and Gautam, Ankur and Raghava, Gajendra PS},
  journal={Biology direct},
  volume={8},
  number={1},
  pages={1--15},
  year={2013},
  publisher={BioMed Central}
}

@article{kadam2021antibody,
  doi={10.3389/fbinf.2021.709951},
  title={Antibody Class (es) Predictor for Epitopes (AbCPE): A Multi-Label Classification Algorithm},
  author={Kadam, Kiran and Peerzada, Noor and Karbhal, Rajiv and Sawant, Sangeeta and Valadi, Jayaraman and Kulkarni-Kale, Urmila},
  journal={Frontiers in Bioinformatics},
  pages={37},
  year={2021},
  publisher={Frontiers}
}

@article{vita2019immune,
  doi={10.1093/nar/gky1006},
  title={The immune epitope database (IEDB): 2018 update},
  author={Vita, Randi and Mahajan, Swapnil and Overton, James A and Dhanda, Sandeep Kumar and Martini, Sheridan and Cantrell, Jason R and Wheeler, Daniel K and Sette, Alessandro and Peters, Bjoern},
  journal={Nucleic acids research},
  volume={47},
  number={D1},
  pages={D339--D343},
  year={2019},
  publisher={Oxford University Press}
}

@article{singh2013improved,
  doi={10.1371/journal.pone.0062216},
  title={Improved method for linear B-cell epitope prediction using antigen’s primary sequence},
  author={Singh, Harinder and Ansari, Hifzur Rahman and Raghava, Gajendra PS},
  journal={PloS one},
  volume={8},
  number={5},
  pages={e62216},
  year={2013},
  publisher={Public Library of Science San Francisco, USA}
}

@article{jespersen2017bepipred,
  doi={10.1093/nar/gkx346},
  title={BepiPred-2.0: improving sequence-based B-cell epitope prediction using conformational epitopes},
  author={Jespersen, Martin Closter and Peters, Bjoern and Nielsen, Morten and Marcatili, Paolo},
  journal={Nucleic acids research},
  volume={45},
  number={W1},
  pages={W24--W29},
  year={2017},
  publisher={Oxford University Press}
}

@article{manavalan2018ibce,
  doi={10.3389/fimmu.2018.01695},
  title={iBCE-EL: a new ensemble learning framework for improved linear B-cell epitope prediction},
  author={Manavalan, Balachandran and Govindaraj, Rajiv Gandhi and Shin, Tae Hwan and Kim, Myeong Ok and Lee, Gwang},
  journal={Frontiers in immunology},
  volume={9},
  pages={1695},
  year={2018},
  publisher={Frontiers Media SA}
}

@article{liu2020deep,
  doi={10.1186/s13040-020-00211-0},
  title={Deep learning methods improve linear B-cell epitope prediction},
  author={Liu, Tao and Shi, Kaiwen and Li, Wuju},
  journal={BioData mining},
  volume={13},
  number={1},
  pages={1--13},
  year={2020},
  publisher={BioMed Central}
}

@article{xu2022netbce,
  doi={10.1016/j.gpb.2022.11.009},
  title={NetBCE: An Interpretable Deep Neural Network for Accurate Prediction of Linear B-Cell Epitopes},
  author={Xu, Haodong and Zhao, Zhongming},
  journal={Genomics, Proteomics \& Bioinformatics},
  year={2022},
  publisher={Elsevier}
}

@article{ofer2021language,
  doi={10.1016/j.csbj.2021.03.022},
  title={The language of proteins: NLP, machine learning \& protein sequences},
  author={Ofer, Dan and Brandes, Nadav and Linial, Michal},
  journal={Computational and Structural Biotechnology Journal},
  volume={19},
  pages={1750--1758},
  year={2021},
  publisher={Elsevier}
}

@article{teufel2022signalp,
  doi={10.1038/s41587-021-01156-3},
  title={SignalP 6.0 predicts all five types of signal peptides using protein language models},
  author={Teufel, Felix and Almagro Armenteros, Jos{\'e} Juan and Johansen, Alexander Rosenberg and G{\'\i}slason, Magn{\'u}s Halld{\'o}r and Pihl, Silas Irby and Tsirigos, Konstantinos D and Winther, Ole and Brunak, S{\o}ren and von Heijne, Gunnar and Nielsen, Henrik},
  journal={Nature biotechnology},
  volume={40},
  number={7},
  pages={1023--1025},
  year={2022},
  publisher={Nature Publishing Group US New York}
}

@article{ma2022identification,
  doi={10.1038/s41587-022-01226-0},
  title={Identification of antimicrobial peptides from the human gut microbiome using deep learning},
  author={Ma, Yue and Guo, Zhengyan and Xia, Binbin and Zhang, Yuwei and Liu, Xiaolin and Yu, Ying and Tang, Na and Tong, Xiaomei and Wang, Min and Ye, Xin and others},
  journal={Nature Biotechnology},
  volume={40},
  number={6},
  pages={921--931},
  year={2022},
  publisher={Nature Publishing Group US New York}
}

@article{thumuluri2022deeploc,
  doi={10.1093/nar/gkac278},
  title={DeepLoc 2.0: multi-label subcellular localization prediction using protein language models},
  author={Thumuluri, Vineet and Almagro Armenteros, Jos{\'e} Juan and Johansen, Alexander Rosenberg and Nielsen, Henrik and Winther, Ole},
  journal={Nucleic acids research},
  volume={50},
  number={W1},
  pages={W228--W234},
  year={2022},
  publisher={Oxford University Press}
}

@article{lecun2015deep,
  doi={10.1038/nature14539},
  title={Deep learning},
  author={LeCun, Yann and Bengio, Yoshua and Hinton, Geoffrey},
  journal={nature},
  volume={521},
  number={7553},
  pages={436--444},
  year={2015},
  publisher={Nature Publishing Group}
}

@article{pearson2013selecting,
  doi={10.1002/0471250953.bi0305s43},
  title={Selecting the right similarity-scoring matrix},
  author={Pearson, William R},
  journal={Current protocols in bioinformatics},
  volume={43},
  number={1},
  pages={3--5},
  year={2013},
  publisher={Wiley Online Library}
}

@article{fu2012cd,
  doi={10.1093/bioinformatics/bts565},
  title={CD-HIT: accelerated for clustering the next-generation sequencing data},
  author={Fu, Limin and Niu, Beifang and Zhu, Zhengwei and Wu, Sitao and Li, Weizhong},
  journal={Bioinformatics},
  volume={28},
  number={23},
  pages={3150--3152},
  year={2012},
  publisher={Oxford University Press}
}

@article{rahman2016inadequate,
  doi={10.1074/jbc.M116.729020},
  title={Inadequate reference datasets biased toward short non-epitopes confound B-cell epitope prediction},
  author={Rahman, Kh Shamsur and Chowdhury, Erfan Ullah and Sachse, Konrad and Kaltenboeck, Bernhard},
  journal={Journal of Biological Chemistry},
  volume={291},
  number={28},
  pages={14585--14599},
  year={2016},
  publisher={ASBMB}
}

@article{ashford2021organism,
  doi={10.1093/bioinformatics/btab536},
  title={Organism-specific training improves performance of linear B-cell epitope prediction},
  author={Ashford, Jodie and Reis-Cunha, Joao and Lobo, Igor and Lobo, Francisco and Campelo, Felipe},
  journal={Bioinformatics},
  volume={37},
  number={24},
  pages={4826--4834},
  year={2021},
  publisher={Oxford University Press}
}

\section*{Ethical Statement}
In this work, we develop a novel computational framework for predicting and classifying B-cell epitopes. The datasets we used are constructed from a publicly available database downloaded from the IEDB website. We do not require ethical approval during the research. Our work does not involve collecting personal data. The major implication will be in the medical domain such as vaccine production and diagnostics development. With that being said, we are aware that some medical applications may need to process personal data. Although the IEDB database does not involve sensitive data such as information on patients, it could be interesting for the community to investigate if the immunology data have a bias in race or gender.

\newpage
\appendix
\section{Derivation of Logit-Adjusted Losses}
In this appendix, we provide a derivation of the logit-adjusted losses for cross-entropy from the perspective of Bayesian decision theory in both the softmax (multiclass) and sigmoid (binary) settings.

Under class imbalance, to minimize the balanced error rate, we use $\mathbb{P}_\mathtt{bal}(y|x)\propto\frac{1}{C}\mathbb{P}(x|y)$, where $\mathbb{P}(x|y)$ is the class-conditional likelihood, rather than the native class posterior $\mathbb{P}(y|x)\propto\mathbb{P}(y)\mathbb{P}(x|y)$. In~\cite{menon2020long}, logit adjustment is proposed to achieves this Bayes-optimal classifier. In the post-hoc logit adjustment, the model is trained with standard cross-entropy and outputs logits $f_y(x)$. Suppose $\mathbb{P}(y|x)\propto\exp(f_y(x))$, using Bayes' theorem, we have:
\begin{equation*}
\begin{aligned}
\arg\max_y\mathbb{P}_\mathtt{bal}(y|x)&=\arg\max_y\mathbb{P}(x|y)=\arg\max_y\frac{\mathbb{P}(y|x)\mathbb{P}(x)}{\mathbb{P}(y)}=\arg\max_y\frac{\exp(f_y(x))}{\pi_y}\\
&=\arg\max_yf_y(x)-\log\pi_y,
\end{aligned}
\end{equation*}
where $\pi_y$ is the empirical class prior.

Alternatively, following~\cite{menon2020long}, we can use the logit adjusted softmax cross-entropy loss:
\begin{equation*}
\ell_\mathtt{softmax}(y,f(x))=-\log\frac{\exp(f_y(x)+\log\pi_y)}{\sum_{i=0}^{C-1}\exp(f_i(x)+\log\pi_i)}.
\end{equation*}
This can be viewed as training a classifier of the form $g_y(x)=f_y(x)+\log\pi_y$, while at test time we predict with $\arg\max_yf_y(x)=\arg\max_yg_y(x)-\log\pi_y$. In other words, the correction based on class priors is baked into the loss during training instead of being applied explicitly in prediction.

We extend the idea to binary classification $y\in\{0,1\}$. Suppose the model outputs logit $f(x)$ and $\mathbb{P}(y=1|x)\propto\exp(f(x))$, we have:
\begin{equation*}
\begin{aligned}
y^*(x)&=\mathbf{1}\left(\mathbb{P}_\mathtt{bal}(y=1|x)>\mathbb{P}_\mathtt{bal}(y=0|x)\right)=\mathbf{1}\left(\mathbb{P}(x|y=1)>\mathbb{P}(x|y=0)\right)\\
&=\mathbf{1}\left(\frac{\mathbb{P}(y=1|x)\mathbb{P}(x)}{\mathbb{P}(y=1)}>\frac{\mathbb{P}(y=0|x)\mathbb{P}(x)}{\mathbb{P}(y=0)}\right)=\mathbf{1}\left(\frac{\exp(f(x))}{\pi}>\frac{\exp(0)}{1-\pi}\right)\\
&=\mathbf{1}\left(f(x)-\log\frac{\pi}{1-\pi}>0\right),
\end{aligned}
\end{equation*}
where $\mathbf{1}(\cdot)$ is the indicator function and $\pi$ is the empirical positive class prior. We therefore shift the logit by the log prior odds.

Similarly, we can define the logit-adjusted sigmoid cross-entropy loss as:
\begin{equation*}
\ell_\mathtt{binary}(y,f(x))=-y\log(\sigma(f(x)+\log\frac{\pi}{1-\pi}))-(1-y)\log(1-\sigma(f(x)+\log\frac{\pi}{1-\pi})).
\end{equation*}
This is equivalent to training a classifier of the form $g(x)=f(x)+\log\frac{\pi}{1-\pi}$, and using the decision rule $\mathbf{1}(f(x)>0)$ at test time.

\end{document}